\documentstyle[preprint,prl,aps]{revtex}

\tightenlines

\input epsf

\title{ Second Cluster Integral and Excluded Volume Effects\\ for the Pion
Gas}

\author{
A.~Kostyuk  $^{1,2}$, M.~Gorenstein $^{1,2}$,
H.~St\"ocker$^1$  and W.~Greiner $^1$
}

\address{
$^1$ Institut f\"ur Theoretische Physik, Universit\"at  Frankfurt,
Germany\\
$^2$ Bogolyubov Institute for Theoretical Physics,
Kyiv, Ukraine
}

\draft

\begin{document}

\maketitle

\begin{abstract}
The quantum mechanical formula for  Mayer's second 
cluster integral for the gas of relativistic particles with hard-core 
interaction is derived. 
The proper pion volume 
calculated with quantum mechanical formula is found to be an order of 
magnitude larger than its classical evaluation.

The second cluster integral for the pion gas is calculated
in quantum mechanical approach with account for both attractive
and hard-core repulsive interactions.
It is shown that,  in the second 
cluster approximation, the repulsive $\pi \pi$-interactions
as well as the finite width of resonances give
important but almost canceling contributions.
In contrast,  
an appreciable deviation from the ideal gas of pions and pion resonances 
is observed beyond the  second cluster approximation
in the framework of the Van der Waals excluded-volume model.
\end{abstract}

%\newpage

\vskip 0.5cm

\section{Introduction}

Thermal models ("fireball") have been popular
for decades (see, e.g. \cite{Sto}) to fit
the data
on multiparticle production in high energy
nucleus--nucleus collisions (see, e.g.  
\cite{Go:99} and references therein). 
The ideal gas (IG) model of noninteracting hadrons and resonances
which has mostly been empoyed to extract
temperature $T$, baryonic chemical potential $\mu_B$, etc.,
from fits to the data is however not adequate for this purpose.
This is among other things because of the following two reasons: 
\begin{itemize}
\item 
The ideal gas model ignores the finite width of the resonances, while
most of them have a
width comparable to or even larger than the typical temperatures of the 
hadron gas $120 \div 180$~MeV. This leads to underestimation of the
attraction
between hadrons.
\item 
The IG model  does not take into account nonresonance interaction
between
hadrons, in particular the repulsion. As a result, 
in the description of the hadron yields data for the AGS and SPS energies 
the IG model 
leads  to artificially large particle number densities,
e.g. $\rho\cong 1.25$~fm$^{-3}$ at $T=185$~MeV and $\mu_B=270$~MeV 
\cite{Go:97},  
which hardly can be consistent with a picture of a gas 
of point-like, noninteracting hadrons. 
\end{itemize}

The procedure of Ref.~\cite{Denisenko87} introduces the Breit--Wigner mass
spectrum 
of resonances. However, this widely used 
procedure works for narrow resonances only.
It was  found that it is insufficient in the realistic case 
\cite{Weinhold97} as it does not take into account 
correlated particle
pairs appearing along with resonances in the hadron gas.
Therefore, the standard procedure underestimates tha attractive part
of hadron interactions. 

To solve the second problem the procedure, which
allows to take into acount finite particcle volume,
was proposed by Hagedorn and Rafelski 
\cite{Hagedorn:1980kb}. The excluded-volume Van der Waals equation of 
state was derived in Ref.\cite{Gor-VdW} and used by several authors 
(see, e.g. \cite{Go:99,Go:97,Venugopalan92} and references therein). Recently,
this procedure was generalized to multicomponent \cite{Gorenstein:1999ce} 
and relativistic particle systems \cite{Bugaev:2000wz}. Still, the proper 
particle volume was so far calculated by {\it classical}
statistical mechanics formulae. 

Our aim is to calculate Mayer's cluster integrals
(CIs) for 
the hadron gas from the available data on 
the hadron scatterings using correct {\it quantum} formulae
and use them for fixing the parameters 
of the Van der Waals excluded volume model. 
In the present paper a first step in this direction is made,
namely we calculate  the second cluster integral (CI) in the case of 
a pure pion gas for a wide temperature range and consider the contribution 
of the repulsive part of the $\pi\pi$ interactions into the CI as an
excluded volume of the Van der Waals model.

% Similar problem was recently studied \cite{Dobado99} 
% in the framework of the 
% chiral perturbation theory. Our approach is based on the best  
% fits to the {\it experimental} data on the $\pi \pi$ scattering, 
% available in the literature.    

The article is organized as follows:
in section II we derive the formula for the second cluster
integral taking into account relativistic effects as well as the
isospin of the pion.
Section III is devoted to the hard-core repulsion
at the quantum level. The domain of applicability of the classical
formulae is found.
The resonance attraction will be considered in section IV. 
The conditions which allow to use the narrow resonance 
approximation (NRA) and the Bright--Wigner formula of 
Ref.~\cite{Denisenko87} will be studied.    
In  section V the CI for the pion gas is calculated from
the experimental data on the $\pi \pi$-scattering. The results
are compared with various approximations.
In section VI
the interacting pion gas is studied in the framework of the 
excluded-volume Van der Waals approach.
The conclusions and a discussion 
of the results are given in section VII.

\section{General formulae}

Quantum mechanical formula for a calculation of Mayer's second cluster
integral 
$b_2$ in the case of nonrelativistic 
zero isospin $I_0=0$ 
particles was 
considered in Ref.~\cite{Beth37} (see also \cite{Huang}). 
The pions, however, have nonzero isospin $I_0=1$ and the
temperature of interest 
can be comparable to or larger than the pion mass. Therefore,
for an adequate description of the pion gas a generalization of the 
formulae given in Refs.~\cite{Beth37,Huang} for relativistic particles
carrying nonzero isospin is needed. 

We start from the
canonical partition function for N identical particles in the volume
$V$ at the temperature $T$ 
\begin{equation} \label{partN}
  Z(V,T,N)=\int d^{3N}r \sum_\alpha 
  \Psi^*_\alpha ({\bf r}_1, {\bf r}_2, \dots {\bf r}_N)
  \exp \left( - \frac{H}{T} \right)
  \Psi_\alpha ({\bf r}_1, {\bf r}_2, \dots {\bf r}_N),  
\end{equation}
where $H$ is the Hamiltonian operator and $\{ \Psi_\alpha  \}$
is a complete set of orthonormal wave functions in co-ordinate
%(${\bf r}_i$,  $i=1, \dots , N$) 
representation.
With the notations 
\begin{equation} \label{WN}
W_N({\bf r}_1, {\bf r}_2, \dots {\bf r}_N)~ \equiv ~ 
N! \sum_\alpha 
  \Psi^*_\alpha ({\bf r}_1, {\bf r}_2, \dots {\bf r}_N)
  \exp \left( - \frac{H}{T} \right)
  \Psi_\alpha ({\bf r}_1, {\bf r}_2, \dots {\bf r}_N)     
\end{equation}
one gets
\begin{equation} \label{partNW}
  Z(V,T,N)~=~ \frac{1}{N!}  \int d^{3N}r 
  W_N({\bf r}_1, {\bf r}_2, \dots {\bf r}_N)~.
\end{equation}
The function $W_1({\bf r}_1)$ can be calculated in the 
thermodynamical limit $V \rightarrow \infty$
\begin{eqnarray}\label{W1}
W_N({\bf r}_1)
&=&
\sum_{{\bf p},t_I}
\frac{e^{-i ({\bf p},{\bf r}_1)}}{\sqrt{V}}
\exp \left( - \frac{H}{T} \right)
\frac{e^{i ({\bf p},{\bf r}_1)}}{\sqrt{V}} \\
&=&
(2 I_0 + 1) \int \frac{d^3p}{(2\pi)^3}  
\exp \left( - \frac{\sqrt{{\bf p}^2+m^2}}{T} \right)
=
g \phi(T;m), \nonumber
\end{eqnarray}
where $I_0$ and $m$ are, respectively, the particle isospin and mass 
($t_I=-I_0,...,+I_0$ is the isospin projection), $g \equiv (2 I_0 + 1)$ is
the isospin
degeneration factor\footnote{
If not only the isospin but
also the spin has a nonzero value, $J_0$, the 
degeneration factor  
has the form $g \equiv (2 I_0 + 1) (2 J_0 + 1)$. 
}  
and $\phi(T;m)$ can be expressed via $K_{2}$ modified Bessel function 
\begin{equation}\label{phi}
\phi(T;m)~ =~ \frac{1}{2 \pi^2} \int_0^{\infty}p^2 dp~
\exp \left( - \frac{\sqrt{p^{2}+m^{2}}}{T}  \right)~
= ~\frac{m^{2} T}{2 \pi^{2}}~ K_{2}\left( \frac{m}{T} \right)~.
\end{equation}
The asymptotics of $\phi(T;m)$ in the nonrelativistic,
$m>>T$, and ultra-relativistic, $m<<T$, limits are 
\begin{equation} \label{phias}
\phi(T;m)~\simeq~\left\{
\begin{array}{ll}
\left( \frac{m T}{2 \pi} \right)^{3/2}~ \exp(-m/T)~, &
m>>T \\
& \\
\frac{T}{\pi^2} \left( T^2 - \frac{m^2}{4} \right)~, & 
m<<T
\end{array}
\right.
\end{equation}

Following Refs.~\cite{Huang,Kahn38} we introduce the functions 
$U_l ({\bf r}_1, {\bf r}_2, \dots {\bf r}_l)$:
\begin{eqnarray} 
W_1({\bf r}_1) &=& U_1({\bf r}_1) \nonumber \\
W_2({\bf r}_1,{\bf r}_2) &=& 
U_1({\bf r}_1) U_1({\bf r}_2) + U_2({\bf r}_1,{\bf r}_2)
\label{Wl} \\ 
W_3({\bf r}_1,{\bf r}_2,{\bf r}_3) &=& 
U_1({\bf r}_1) U_1({\bf r}_2) U_1({\bf r}_3)  
+ U_1({\bf r}_1) U_2({\bf r}_2,{\bf r}_3)   \nonumber \\ & &
+ U_1({\bf r}_2) U_2({\bf r}_3,{\bf r}_1)
+ U_1({\bf r}_3) U_2({\bf r}_1,{\bf r}_2)
+ U_3({\bf r}_1,{\bf r}_2,{\bf r}_3)
\nonumber \\
\mbox{etc.}  \nonumber 
\end{eqnarray}
and define Mayer's CIs\footnote{
The normalizations of the CIs in Eq.~(\ref{bl}) are different from that
of Ref.\cite{Huang} and correspond to the definition 
used in \cite{Mayer}. The CIs (\ref{bl}) have dimensionality
$[\mbox{volume}]^{l-1}$,  while in Ref.\cite{Huang} CIs are
dimensionless.}
\begin{equation}\label{bl}
 b_l(V,T)~ =~ \frac{1}{l! V [ g \phi(T;m) ]^l}
\int d^{3l}r ~ U_l({\bf r}_1, {\bf r}_2, \dots {\bf r}_l).
\end{equation}
It is easy to see that 
\begin{equation}\label{b1}
 b_1~\equiv~1~.
\end{equation}

Substituting the expression (\ref{Wl}) for the function $W_2$ into 
Eq.~(\ref{partNW}) one gets the two-particle partition function expressed
via the CIs
\begin{equation} \label{part2b}
  Z(V,T,2)~=~ g^2 \left[
	\frac{1}{2} (b_1(V,T) \phi(T;m) V )^2 + 
	b_2(V,T) \phi^2(T;m) V ~  \right].
\end{equation}
For arbitrary $N$ the expression of the partition function via 
$b_l$ reads \cite{Huang}
\begin{equation} \label{partNb}
 Z(V,T,N) ~=~ \sum_{ \{  m_l \} } \prod_{l=1}^N 
 \frac{1}{m_l !} (b_l(V,T) \left[ g \phi(T;m) \right]^l  V )^{m_l}~,
\end{equation}
where the sum runs over all sets of nonnegative integer numbers 
$ \{  m_l \} $ satisfying
the condition 
\begin{equation} \label{cond}
\sum_{l=1}^N l m_l ~ =~ N~.
\end{equation}
Introducing the absolute activity \cite{Mayer}, which in
the relativistic case takes the form
\begin{equation} \label{ztild}
 z~\equiv~ g \phi(T;m)   
 \exp\left( \frac{\mu}{T}  \right) ~,
\end{equation}
where $\mu$ is the chemical potential,
the grand canonical partition function has the form
\begin{equation} \label{grand}
 {\cal Z}(V,T,\mu) ~=~ \sum_{N=1}^{\infty} 
 \exp\left( \frac{\mu N}{T}  \right) Z(V,T,N)~ =~
 \exp\left( V  \sum_{l=1}^{\infty} b_l(V,T) z^l \right)~.
\end{equation}

From Eq.~(\ref{grand}) one can find the {\it cluster} expansion of
the pressure and the particle density:
\begin{eqnarray}
\label{pclust}
p(T,\mu)~ &=&~ T \lim_{V \rightarrow \infty}
\frac{\log {\cal Z}(V,T,\mu)}{V}~ =~
T \sum_{l=1}^{\infty} b_l(T) z^l~,  \\
\label{nclust}
n(T,\mu)~ &=&~ \lim_{V \rightarrow \infty} 
\frac{T}{V} 
\frac{\partial \log {\cal Z}(V,T,\mu)}{\partial \mu}~ =~
\sum_{l=1}^{\infty} l b_l(T) z^l~,
\end{eqnarray}
where
\begin{equation}
b_l(T) \equiv \lim_{V \rightarrow \infty} b_l(V,T).
\end{equation}

Substituting the particle density (\ref{nclust}) into the {\it virial}
expansion\footnote{
Sometimes the term 'virial expansion' is used  
in place of 'cluster expansion' \cite{Dashen69,Dobado99}. We prefer to use 
the standard terminology \cite{Huang,Mayer,Greiner}: 'cluster expansion' for
the expansion in powers of the activity and 'virial expansion' for
that in powers of the particle density. 
} 
for the pressure
\begin{equation} \label{virexpan}
p(T, n) = T \sum_{i=1}^{\infty} a_i n^i
\end{equation}
and equating the coefficient of each power of $z$ with
Eq.(\ref{pclust}) one obtains the following 
expressions for the virial coefficients \cite{Huang}
in terms of the CIs:  
\begin{eqnarray}
a_1 &=& 1 \nonumber \\
a_2 &=& - b_2 \nonumber  \\
a_3 &=& 4 b_2^2 - 2 b_3 
\label{vircoef} \\
a_4 &=& - 20 b_2^3 + 18 b_2 b_3 - 3 b_4  \nonumber \\
\mbox{\dots}	\nonumber			
\end{eqnarray}
  
Let us represent the CIs as a sum of two terms:
\begin{equation} \label{blsum}
b_l = b_l^{(0)} + b^{(i)}_l~, \ \ \ \ \ \ \ \ \ 
l > 1,    
\end{equation}
where $b^{(0)}_l$ are the CIs for the IG and $b^{(i)}_l$
appear due to the particle interaction.
In the classical (Boltzmann) gas one obtains $b^{(0)}_l = 0$ for all
$l > 1$. In the quantum case, $b^{(0)}_l$ are nonzero due to Bose
(Fermi) effects and can be easily 
found for arbitrary $l$. For noninteracting particles the logarithm of 
the expression (\ref{grand}) should coincide with the well-known expression
for the logarithm of the ideal gas grand canonical partition function
\begin{equation} \label{grandPG}
 \log {\cal Z}^{(0)}(V,T,\mu)~ =~ \pm g V 
 \int \frac{d^3 p}{(2\pi)^3} \log \left[1 \pm 
 \exp \left( \frac{\mu - \sqrt{{\bf p}^2+m^2}}{T} \right) 
\right]
\end{equation}
(the upper (lower) sign corresponds to Fermi-Dirac (Bose-Einstein)
statistics). 
One can expand the logarithm in the integrand and perform the integration
\begin{eqnarray} \label{grandPGexp}
  \log {\cal Z}^{(0)}(V,T,\mu)~ &=&~ 
  g V \sum_{l=1}^{\infty} 
  \frac{(\mp 1)^{l+1}}{l}
  \int \frac{d^3 p}{(2\pi)^3}
  \exp \left( \frac{ 
   l \left( \mu - \sqrt{{\bf p}^2+m^2} \right)
  }{T} \right)  \nonumber \\ 
  &=&~
  g V \sum_{l=1}^{\infty}
  \frac{(\mp 1)^{l+1}}{l}
  \exp \left( \frac{l \mu}{T} \right)
  \phi(T/l;m) \ .
\end{eqnarray}
Comparing the last expression with Eq.~(\ref{grand}) gives
\begin{equation} \label{blpg}
  b_l^{(0)}~ =~ \frac{(\mp 1)^{l+1}}{l g^{l-1}}~ 
  \frac{\phi(T/l;m)}{ [ \phi(T;m) ]^l}.
\end{equation}

In the nonrelativistic limit Eq.(\ref{blpg}) is reduced to
\begin{equation} \label{blnonrel}
  b^{(0)}_l~ =~  (\mp 1)^{l+1} l^{-5/2}
  \left[ \frac{\lambda^3}{g} \right]^{l-1}~, 
\end{equation}
where $\lambda$ is the thermal wave length
\begin{equation} \label{lambda}
\lambda~ =~  \sqrt{ \frac{2 \pi}{m T} }~ .
\end{equation}
The expression (\ref{blnonrel}) coincides for $I_0=0$
with the corresponding formulae of Ref.~\cite{Huang}
(up to the dimensional factor $\lambda^{3(l-1)}$,
because of different normalization in Eq.~(\ref{bl})).

Using Eqs.~(\ref{part2b}) and (\ref{blsum}) one can express 
$b^{(i)}_2$ via differences of the two-particle partition 
functions for real and ideal gases:
\begin{equation} \label{b2Z}
 b^{(i)}_2 ~=~ \frac{Z(V,T,2) - Z^{(0)}(V,T,2)}
 {V [g \phi(T;m) ]^2}
\end{equation}

Let us calculate $Z(V,T,2)$. A complete set of
the orthonormal state vectors
in the two particle system can be constructed from the following wave 
functions
\begin{equation} \label{wf}
   | \alpha \rangle ~\equiv ~  |{\bf P} ,\tilde{\alpha} \rangle
 ~ =~ \frac{e^{i({\bf P},{\bf R})}}{\sqrt{V}} | \tilde{\alpha} \rangle~, 
\end{equation}
where ${\bf P}$ is the total momentum of the system,  
${\bf R}$ is the radius-vector of its center of mass,
and $ | \tilde{\alpha} \rangle $ form a complete set of orthonormal
state vectors of the system in the center of mass frame (c.m.f.)  
satisfying the Schr\"odinger equation  
\begin{equation} \label{sroedcms}
   H \ | \tilde{\alpha} \rangle ~ = ~ 
   \varepsilon_{\tilde{\alpha}} \ | \tilde{\alpha} \rangle
\end{equation}
with the normalization condition
\begin{equation} \label{normcms}
   \langle \tilde{\alpha}^{\prime} | \tilde{\alpha} \rangle ~=~
   \delta_{\tilde{\alpha}^{\prime}  \tilde{\alpha}}~.
\end{equation}
The wave function (\ref{wf}) thus satisfies the following equations
\begin{equation} \label{sroed}
   H \ | {\bf P} ,\tilde{\alpha} \rangle = 
   \sqrt{ {\bf P}^2 + \varepsilon_{\tilde{\alpha}}^2 } \  
   | {\bf P} ,\tilde{\alpha} \rangle~,
\end{equation}
\begin{equation} \label{norm}
   \langle {\bf P}^{\prime} , \tilde{\alpha}^{\prime} | 
   \tilde{ {\bf P} , \alpha} \rangle~ =~
   \delta_{\tilde{\alpha}^{\prime}  \tilde{\alpha}}
   \delta_{ {\bf P}^{\prime}{\bf P} } \ .
\end{equation}

The expression for $Z(V,T,2)$ in terms of the introduced wave functions
has the form 
\begin{equation} \label{Z2P}
Z(V,T,2)~ =~ 
\sum_{{\bf P} ,\tilde{\alpha}}
\langle {\bf P} ,\tilde{\alpha} |
\exp \left( - \frac{H}{T} \right)
|{\bf P} ,\tilde{\alpha} \rangle 
~=~
\sum_{{\bf P} ,\tilde{\alpha}}
\exp \left( - \frac{
\sqrt{ {\bf P}^2 + \varepsilon_{\tilde{\alpha}}^2 }
}{T} \right)~.
\end{equation}
In the thermodynamical limit $V \rightarrow \infty$ the summation 
over ${\bf P}$ can be replaced by the integration and one finds
\begin{equation} \label{Z2K} 
 Z(V,T,2) ~=~ 
 \sum_{\tilde{\alpha}} 
 V  \int \frac{d^3 P}{(2\pi)^3}
\exp \left( - \frac{
\sqrt{ {\bf P}^2 + \varepsilon_{\tilde{\alpha}}^2 }
}{T} \right) ~
=~
V \sum_{\tilde{\alpha}}
\phi(T; \varepsilon_{\tilde{\alpha}} )~.
\end{equation}

The states of two spinless particles in their c.m.f.
can be enumerated by the following quantum numbers:
the radial momentum $q$ (or, alternatively, the energy 
in c.m.f. $\varepsilon(q)=\sqrt{q^2+m^2})$,
the orbital angular momentum $L$, its projection $m_L$,
the total isospin $I$ and its projection $t_I$, e.g.
$\tilde{\alpha} = (q,L,m_L,I,t_I)$. Eq.(\ref{Z2K}) can be
rewritten explicitly \footnote{We assume that the
particles do not form bound states with energy 
$\varepsilon < 2m $. 
}  
\begin{equation} \label{Z2Kexp} 
 Z(V,T,2)~ =~ V \sum_{I=0}^{2 I_0}  \sum_{t_I=-I}^{I}
 {\sum_{L}}^\prime \sum_{m_L=-L}^{L}
 \int_0^\infty d q g_{L m_L I t_I}(q) \phi(T; \varepsilon(q))~,
\end{equation}
where $g_{L m_L I t_I}(q)$ is the density of states with the  
given set of quantum numbers.
The sum ${\sum}^\prime$ extends over only those values of $L$ 
that satisfy the symmetry properties of the wave function.
For spinless bosons it takes even value if the isospin part of
the wave function is symmetric and odd values if it is antisymmetric.
In the case of integer isospin particles like pions this means
\footnote{In the
case of half-odd isospin scalar bosons we would have the opposite rule:
even $L$ for odd $I$ and odd $L$ for even $I$. }
\begin{equation} \label{Lval} 
 L~ = ~\left\{
   \begin{array}{ll}
   0,2,4,6, \dots & \mbox{for even } I  \\
   1,3,5,7, \dots & \mbox{for odd  } I  \\ 
   \end{array} 
 \right.
, 
\end{equation}
 Substituting Eq.(\ref{Z2Kexp})
into (\ref{b2Z}) one gets the following expression for
the second CI
\begin{equation} \label{b2g}
  b^{(i)}_2 ~=~ \frac{1}{[ g \phi(T;m) ]^2}
  \sum_{I=0}^{2 I_0}  \sum_{t_I=-I}^{I}
 {\sum_{L}}^\prime \sum_{m_L=-L}^{L}
 \int_0^\infty d q 
 (g_{L m_L I t_I}(q) - g_{L m_L I t_I}^{(0)}(q) )  
 \phi(T; \varepsilon(q))~,
\end{equation}
where $g_{L m_L I t_I}^{(0)}$ is the state density for the IG.
The difference $g_{L m_L I t_I}(q) - g_{L m_L I t_I}^{(0)}(q)$
in the thermodynamical limit can be expressed via phase shifts
of two-particle scattering  $\delta_{L m_L I t_I}(q)$ 
\cite{Beth37,Huang}:
\begin{equation} \label{delta}
   g_{L m_L I t_I}(q) - g_{L m_L I t_I}^{(0)}(q)~ =~
   \frac{1}{\pi} \frac{d \delta_{L m_L I t_I}(q)}{d q}~.
\end{equation}
Using the expression for $\phi(T;m)$ one gets
\begin{equation} \label{b2delta}
  b^{(i)}_2 ~=~ \frac{2 \pi}{ m^4 T [ g K_2(m/T) ]^2}
  \sum_{I=0}^{2 I_0}  \sum_{t_I=-I}^{I}
 {\sum_{L}}^\prime \sum_{m_L=-L}^{L}
 \int_0^\infty d q  \varepsilon^2(q)
 \frac{d \delta_{L m_L I t_I}(q)}{d q}   
 K_2(\varepsilon(q)/T)~.
\end{equation}
In the case of hadron gas the phase shift does not depend on 
the angular momentum projection $m_L$ (no external fields)
and the isospin projection $t_I$ (if only strong interactions are
taken into account). This simplifies the last formula:
\begin{equation} \label{b2ddelt}
  b^{(i)}_2~ =~ \frac{2 \pi}{ m^4 T [ g K_2(m/T) ]^2}
  \sum_{I=0}^{2 I_0}  
 {\sum_{L}}^\prime (2I+1) (2L+1)
 \int_0^\infty d q  \varepsilon^2(q) 
 \frac{d \delta_{L I}(q)}{d q}   
 K_2 \left( \frac{\varepsilon(q)}{T} \right)~.
\end{equation}
Performing a partial integration and taking into account the properties
of the Bessel functions, one gets
\begin{equation} \label{b2delt}
  b^{(i)}_2~ =~ \frac{2 \pi}{ m^4 T^2 [ g K_2(m/T) ]^2}
  \sum_{I=0}^{2 I_0}  
 {\sum_{L}}^\prime (2I+1) (2L+1)
 \int_{2 m_{\pi}}^\infty d \varepsilon  \varepsilon^2
 \delta_{L,I}(\varepsilon)   
 K_1 \left( \frac{\varepsilon}{T} \right)~.
\end{equation}

In the nonrelativistic limit the formula (\ref{b2ddelt}) is reduced 
to
\begin{equation} \label{b2ddeltnr}
  b^{(i)}_2 = \frac{2 \sqrt{2}}{\pi g^2} \lambda^3
  \sum_{I=0}^{2 I_0}  
 {\sum_{L}}^\prime (2I+1) (2L+1)
 \int_0^\infty d q  
 \frac{d \delta_{L I}(q)}{d q}   
 \exp \left( - \frac{q^2}{m T} \right),
\end{equation}
which again at $I_0=0$ coincides with the corresponding formulae
of Refs.~\cite{Beth37,Huang}, up to the factor $\lambda^3$.

\section{Hard core repulsion}

The hard core repulsion plays an important role in the phenomenological 
description of the $\pi \pi$-scattering: the phase shift data 
for the isospin state $I=2$ can be successfully described assuming 
hard core repulsion between two particles \cite{Ishida97}. The best fit 
of the phase shift in $S_0$-state ($I=L=0$) 
can be obtained by assuming hard core repulsion 
in addition to resonance attraction \cite{Ishida96,Ishida99}.
Hence we start our analysis from applying  
Eqs.(\ref{b2ddelt}--\ref{b2ddeltnr}) to hard sphere model.\footnote{
Relativistic consideration of a hard sphere model by no means can
be consistent. Still, as far as this model describes experimental 
data on low energy $\pi \pi$-scattering, we find it phenomenologically
satisfactory.}

The radial part of the wave function in the c.m.f. of two particles 
interacting by hard core potential in the state with orbital momentum
$L$ and radial momentum $q$ can be represented in the following way
\begin{equation} \label{hsrwf} 
 \phi(r) = \left\{
   \begin{array}{ll}
   0  & \mbox{for } r \le r_0  \\
   C (\cos \delta_L \  j_L(qr) + \sin \delta_L \  y_L(qr)) & 
   \mbox{for } r > r_0  \\ 
   \end{array} 
 \right.
 ,
\end{equation}
where $r$ is the distance between particle centers, $r_0$ is the minimal
admitted value for $r$ (that is the doubled radius of 
the particle considered as a hard sphere), $j_n(z)$ and $y_n(z)$
are spherical Bessel functions, $C$ is the normalization constant
and $\delta_L$ is fixed by the condition 
\begin{equation} \label{hsbc} 
   \cos \delta_L \  j_L(qr_0) + \sin \delta_L \  y_L(qr_0) = 0 \ .  
\end{equation}
Using asymptotic properties of the spherical Bessel functions, it 
is easy to see that $\delta_L$ has a meaning of the phase shift
describing scattering of two hard spheres: 
\begin{equation} \label{hsasym} 
 \phi(r) \simeq   
 \frac{\sin \left( qr -\frac{l \pi}{2} + \delta_L \right)}{q r},
 \mbox{\ \ } 
 r \rightarrow \infty  \ .
\end{equation}

The derivative of the phase shift is found to be
\begin{equation} \label{hsder}
 \frac{d \delta_L}{d q} = \frac{d }{d q} \arctan 
 \left(
   \frac{j_L(qr_0)}{y_L(qr_0)}
 \right) = 
 - \frac{r_0}{ (qr_0)^2 [ j_L^2(qr_0) + y_L^2(qr_0) ] } \ ,
\end{equation}
where the formula for the Wronskian \cite{Abramowitz} 
\begin{equation} \label{Wronsk}
  W(j_L(z),y_L(z)) = z^{-2}  
\end{equation}
has been used.

Let us introduce the functions
\begin{equation}\label{kappa}
\kappa^{\pm}(z) = {\sum_{L}}^\prime
\frac{1}{ z^2 [ j_L^2(z) + y_L^2(z) ] } \ ,
\end{equation}
where the sum ${\sum_{L}}^\prime$ runs over either even 
(superscript `$+$') or odd (superscript `$-$') nonnegative numbers.

Expanding $\kappa^{\pm}$ around zero, one gets
\begin{eqnarray}\label{kappa0}
\kappa^{+}(z) &\simeq& 1+ \frac{5}{9} z^4 + O(z^6) \ ,  \\
\kappa^{-}(z)  &\simeq& 3 z^2 - 3 z^4 + 
\frac{682}{225} z^6 + O(z^8) \ .
\end{eqnarray}

It has been checked numerically that the asymptotic behavior of
$\kappa^{\pm}(z)$ at large $z$ with a high accuracy can be 
presented by the formula
\begin{equation}\label{kappainf}
\kappa^{\pm}(z) \simeq  
\frac{1}{3}z^2 + \frac{\pi}{4}z + \frac{1}{3} + O(z^{-1}) \ .
\end{equation}

The CI  (\ref{b2ddelt}) for the case of hard sphere model can be
represented in the form\footnote{
In the simplest version of the hard sphere model when
the core radius $r_0$ does not depend on $I$, the formula 
(\ref{b2b2pm}) could be reduced to $b^{(i)}_2 = b^{\pm}_2(r_0,T,m)$. 
However, in the case of realistic $\pi \pi$ 
interaction every isospin state has its own $r_0(I)$ 
\cite{Ishida96,Ishida97,Ishida99}.  
}
\begin{equation} \label{b2b2pm}
  b^{(i)}_2 = \frac{1}{g^2}
	\sum_{I=0}^{2 I_0} (2I+1) 
  b^{\pm}_2(r_0,T,m) \ ,
\end{equation}
where superscript `$+$'(`$-$') corresponds to even (odd)
values\footnote{Again, for the case of half-odd isospin particles 
the opposite rule would be valid.}  of $I$  and
$b_2^{\pm}(r_0)$ is expressed via the function 
$\kappa^{\pm}(z)$
\begin{equation} \label{b2evod}
   b^{\pm}_2 = - \frac{2 \pi r_0}{ m^4 T [ K_2(m/T) ]^2}
 \int_0^\infty d q  [ \varepsilon(q) ]^2
 \kappa^{\pm}(q r_0)
 K_2 \left( \frac{\varepsilon(q)}{T} \right).
\end{equation}
In the nonrelativistic approximation the expression for 
$b_2^{\pm}(r_0)$ is reduced to
\begin{equation} \label{b2evodnr}
  b_2^{\pm}(r_0,T,m) 
  = - \frac{2 \sqrt{2}}{\pi} \lambda^3 r_0
 \int_0^\infty d q  
 \kappa^{\pm}(q r_0)    
 \exp \left( - \frac{q^2}{m T} \right).
\end{equation}

Substituting Eq.~(\ref{kappa0}) into the last formula one gets the
nonrelativistic expression for $b_2^{\pm} (r_0,T,m)$ at 
small $r_0$ ($r_0 << \lambda$)
\begin{eqnarray}\label{b2r0sme}
b_2^{+} (r_0,T,m) &\simeq& 
-2 \lambda^2 r_0 \left( 1 
+ \frac{5}{3} \pi^2 (r_0/\lambda)^4
+ O \left[ (r_0/\lambda)^6 \right] 
\right) , \\ \label{b2r0smo}
b_2^{-} (r_0,T,m) &\simeq& 
-6 \pi r_0^3  \left( 1
- 3 \pi (r_0/\lambda)^2
+ \frac{682}{45} \pi^2 (r_0/\lambda)^4
+ O \left[ (r_0/\lambda)^6 \right]
\right) , 
\end{eqnarray}
in the opposite case $r_0 >> \lambda$ the nonrelativistic expression for 
$b_2^{\pm} (r_0,T,m)$ can be obtained using Eq.(\ref{kappainf}):   
\begin{equation} \label{b2r0la}
b_2^{\pm} (r_0,T,m) = 
- \frac{2}{3} \pi r_0^3 
\left( 1
 + \frac{3 \sqrt{2}}{4} \frac{\lambda}{r_0}
+ \frac{3}{2 \pi} \frac{\lambda^2}{r_0^2}
+  O \left[ \left( \frac{\lambda}{r_0}  \right)^3 \right]  
\right)
\end{equation}

\begin{figure}[t]
\begin{center}
\vfill
\leavevmode
\epsfysize=20cm \epsfbox{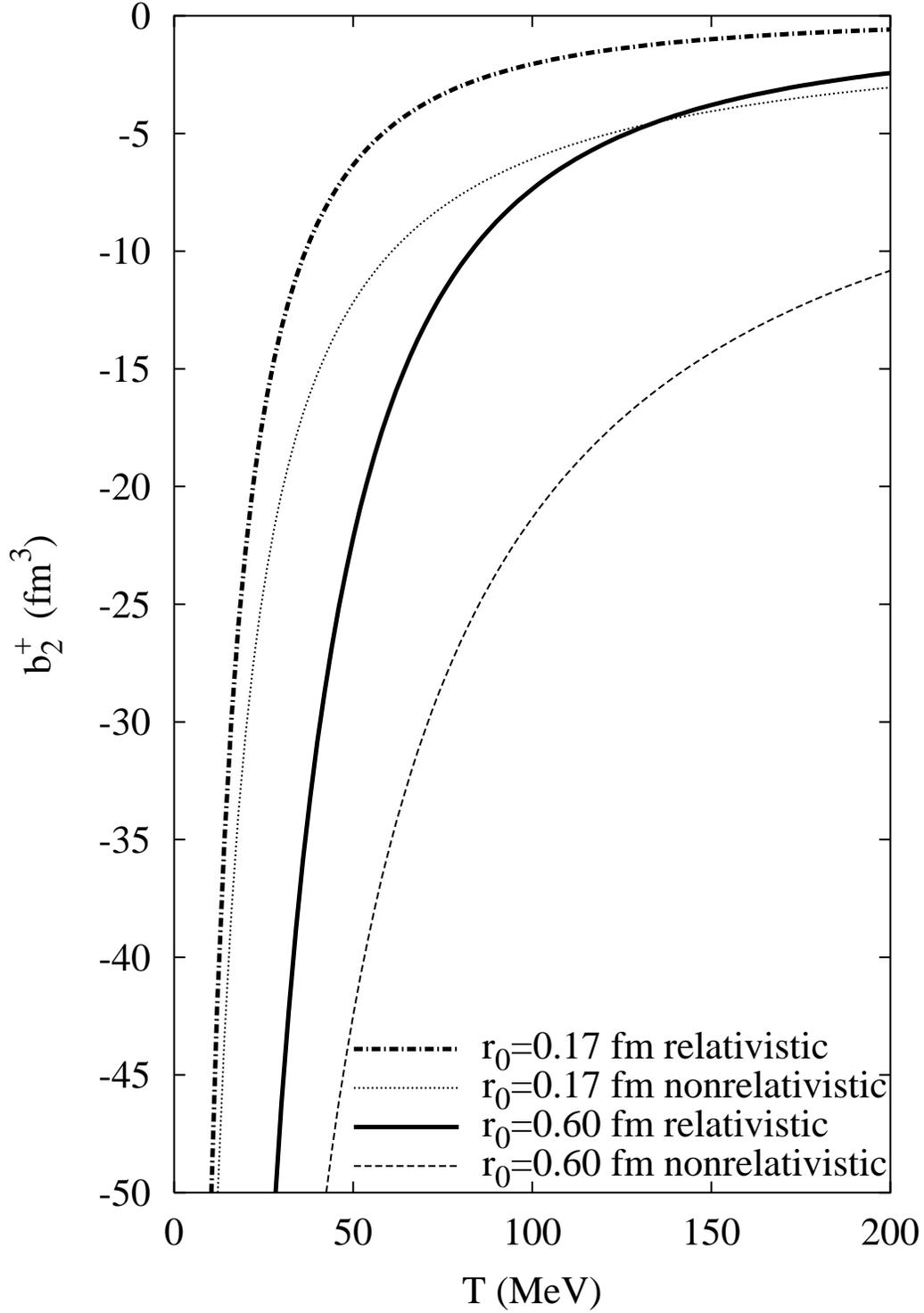}
\vfill
\mbox{}\\
\caption{The dependence of the CI $b_2^{+} (r_0,T,m)$ on the
temperature in the hard sphere model. The formula (\ref{b2evod}) is
used for the relativistic calculations. The results are compared to 
the nonrelativistic approximation (\ref{b2evodnr}).
\label{frepT}
}
\end{center}
\end{figure}

The pion thermal wave length at the temperature $T=50 \div 200$ MeV 
ranges between $3 \div 6$ fm. From 
Eq.~(\ref{b2r0la}) one sees that the classical formula \cite{Greiner}
\begin{equation} \label{class} 
b_2^{\pm} (r_0,T,m) \approx - \frac{2}{3} \pi r_0^3
\end{equation}
(the particle volume multiplied by $4$)
would give a reasonable approximation only at unrealistically 
large hard core radius $r_0 \geq 50$ fm. 
The hard core radii found from the $\pi \pi$-scattering
are much smaller: $r_0 = 0.60$ fm in the $S_0$
state  \cite{Ishida99} and $r_0 = 0.17$ fm 
at $I=2$ \cite{Ishida97}. (No evidence of hard core repulsion was
found in $P_1$-state (I=L=1)). In this case the value of 
$b_2^{+} (r_0,T,m)$ can be estimated from formula 
(\ref{b2r0sme}). The results are presented in Fig.\ref{frepT}.  
It is seen that in contrast to the classical case the quantum treatment 
leads to a rather strong dependence of the second CI on the temperature
(approximately proportional to $1/T$) even in the nonrelativistic
approximation.
Numerical calculations show that relativistic effects make this 
dependence even stronger and essentially reduce at high temperature 
the CI with respect to its nonrelativistic value (see Fig.~\ref{frepT}).
Both relativistic and nonrelativistic quantum formulae give a much  
($1 \div 2$ orders of magnitude)
larger value than those given by the classical formula 
(\ref{class}): $0.45$ fm${}^3$ and $0.01$ fm${}^3$ for $r_0=0.60$ fm 
$r_0=0.17$ fm, respectively.

As can be seen from Fig.\ref{frepT}, the relativistic effects cannot
be ignored even at relatively low temperatures. Therefore, only the
relativistic formula  (\ref{b2evod}) is used in the following
calculations.

\section{Resonance attraction}

The phase shifts of $\pi \pi$ elastic scattering can be approximately 
described by assuming that the attractive parts of the interaction appear
due 
to the propagation of resonances in the $s$-channel of the reaction
(see \cite{Ishida97,Ishida96,Ishida99,Hyams73} and references therein).

The distinctive feature of the resonance interaction is the rapid growth
of the phase shift by $\pi$ radian in the vicinity of the pion momentum 
$q_r$, which is related to the resonance mass $M_r$:
\begin{equation}\label{qr}
      M_r = 2 \sqrt{q_r^2 + m^2} \ .
\end{equation} 
In the limit $\Gamma_r \rightarrow 0$ ($\Gamma_r$ is the resonance width)
the derivative of the phase shift can be approximated by
the Dirac delta-function:
\begin{equation}\label{pidelta}
  \frac{d \delta_{L,I}(q)}{d q} \approx \pi \delta(q - q_r) \ .
\end{equation}
In this approximation, which we will call the 'narrow resonance
approximation' (NRA), the expression (\ref{b2ddelt}) for $b^{(i)}_2$
yields: 
\begin{equation}\label{b2NRA}
  b^{(i)}_2 \approx \frac{1}{g^2} 
	\sum_r (2 I_r + 1) (2 L_r + 1)
  \frac{\phi(T;M_r)}{[ \phi(T;m) ]^2} \ ,
\end{equation}
where $I_r$, $L_r$ and $M_r$ are, respectively, the resonance's isospin,
spin and mass, with the index $r$ running over all resonances in the 
two-pion system.

Eq.(\ref{b2NRA}) allows to rewrite the expression for the grand 
canonical partition function  (\ref{grand}) in the  following form  
\begin{equation} \label{grandNRA}
 {\cal Z}(V,T,\mu) = 
 \exp\left[ V  \left(
 \sum_{l=1}^{\infty} b^{(0)}_l z^l +
 \sum_r  z_r
 \right) \right],
\end{equation}
where
\begin{equation} \label{actNRA}  
z_r=   \phi(T;M_r)
\exp\left( \frac{2 \mu}{T}  \right) 
\end{equation}
is the absolute activity of the resonance $r$ with degeneration factor    
$g_r = (2 I_r + 1) (2 L_r + 1)$.
The expression (\ref{grandNRA}) is nothing else than the partition
function
for a mixture of ideal gases of pions and two-pion resonances\footnote{
The fact that the resonance gases are classical is an artifact 
of the second cluster approximation. The quantum correction would 
appear from the 4-th and higher cluster integrals for the interacting 
pions.}. This recovers the well known result of Ref.\cite{Dashen69} that
narrow resonances contribute to the partition function as 
an ideal gas of stable particles. 
The quantitative criterion for an applicability of the NRA was found
to be \cite{Dashen69}
\begin{equation} \label{NRA}
 \Gamma_r << T  \ .
\end{equation}
The resonances appearing in $\pi \pi$-scattering do not satisfy this 
criterion: most of them ($\rho(770)$, $f_0(980)$, $f_2(1270)$ , 
$\rho_3(1690)$) 
have widths comparable with a typical temperature of the hadron gas
and the width of $f_0(400-1200)$ (known also as the $\sigma$) is a few 
times larger then the temperature. Therefore, it is necessary to take 
into account the finite width of the resonances.  

% in our further calculations 
% we take into account the finite width of the resonances substituting 
% parametrizations that fit experimental data  for the $\pi \pi$-phase 
% shifts \cite{Ishida96,Ishida99,Hyams73} into the formula (\ref{b2delt}).

The scalar-isoscalar resonances contribution to the $\pi \pi$-phase shift
in the $S_0$ state
can be parametrized in the following way \cite{Ishida96}: 
\begin{equation} \label{res0}
  \tan \delta_{r}(q) = \frac{q}{q_r} \frac{M_r^2}{\varepsilon(q)} 
	\frac{\Gamma_r} {M_r^2 - \varepsilon^2(q)} \ ,
	\ \ \ 
	r=\sigma,\ f_0(980) \ . 
\end{equation}
For parametrization of nonzero (iso-)spin resonances we shall use the 
following formula \cite{Hyams73}
\begin{equation} \label{res}
  \tan \delta_{r}(q) = \left( \frac{q}{q_r} \right)^{2 L_r + 1}  
	\frac{M_r x_r \Gamma_r} {M_r^2 - \varepsilon^2(q)}
	\frac{D_{L_r}(q_r R_r)}{D_{L_r}(q R_r)} \ ,
	\ \ \ \ \ 
	r=\rho(770), \ f_2(1270), \  \rho_3(1690),
\end{equation}
where $x_r$ is the inelasticity, i.e. the decay fraction of the resonance
into two pions, $R_r$ is the so-called interaction radius and the
functions
$D_L(z)$ are given by the formulae
\begin{eqnarray}
D_1(z) &=& 1 + z^2 \nonumber \\
D_2(z) &=& 9 + 3 z^2 + z^4 \\
D_3(z) &=& 225 + 45 z^2 + 6 z^4 + z^6 \ . \nonumber 
\end{eqnarray}
The resonance parameters are given in the Table \ref{respar}.

\begin{figure}[t]
\begin{center}
\vfill
\leavevmode
\epsfysize=20cm \epsfbox{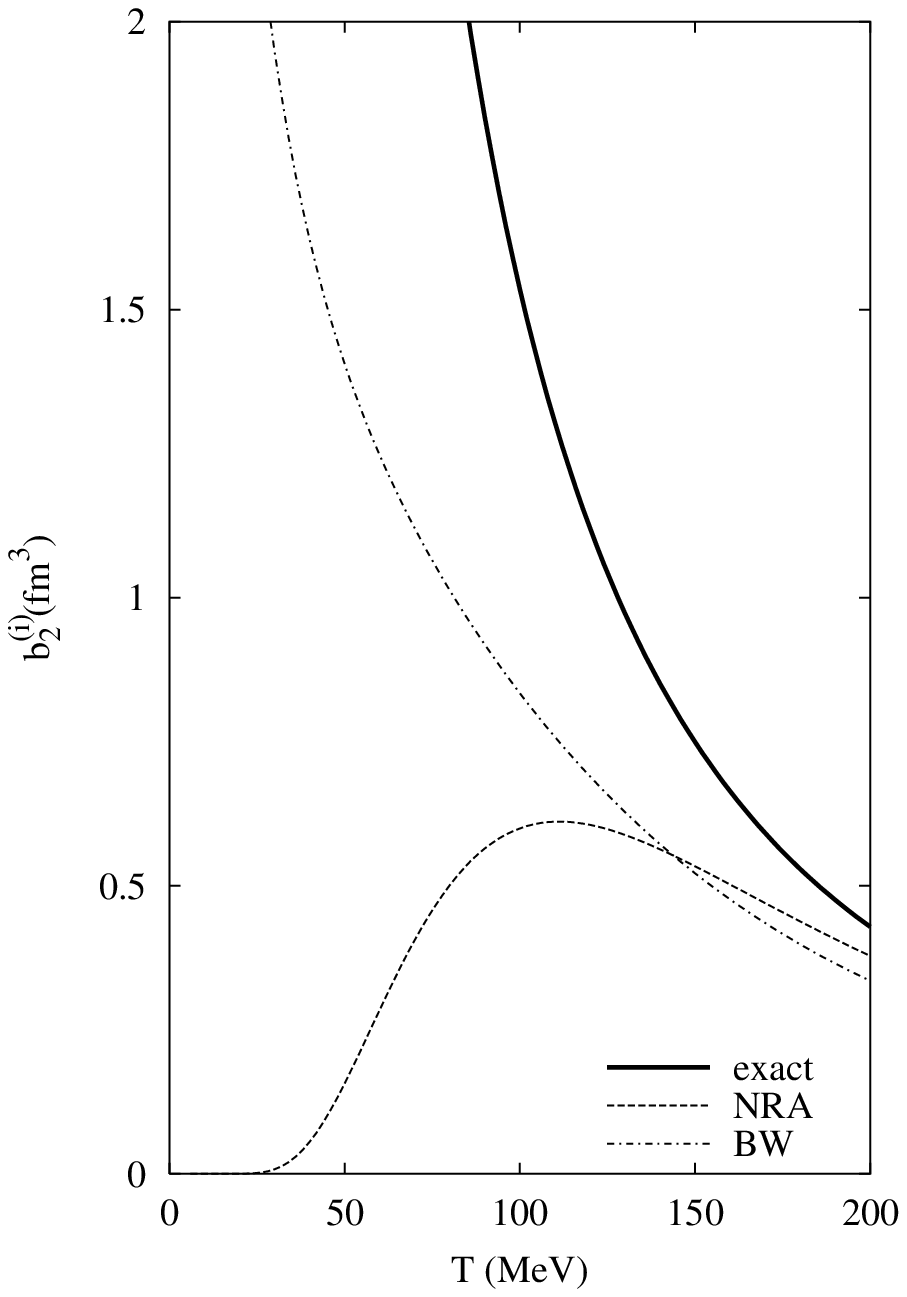}
\vfill
\mbox{}\\
\caption{
The contribution of  $\sigma$-resonance to $b_2^{(i)} (T)$.
The exact value is compared to those calculated in narrow 
resonance approximation (NRA) and using normalized Breit-Wigner (BW)
profile of the resonance (\ref{BW-b2}).
\label{fsigma}
}
\end{center}
\end{figure}

\begin{figure}[t]
\begin{center}
\vfill
\leavevmode
\epsfysize=20cm \epsfbox{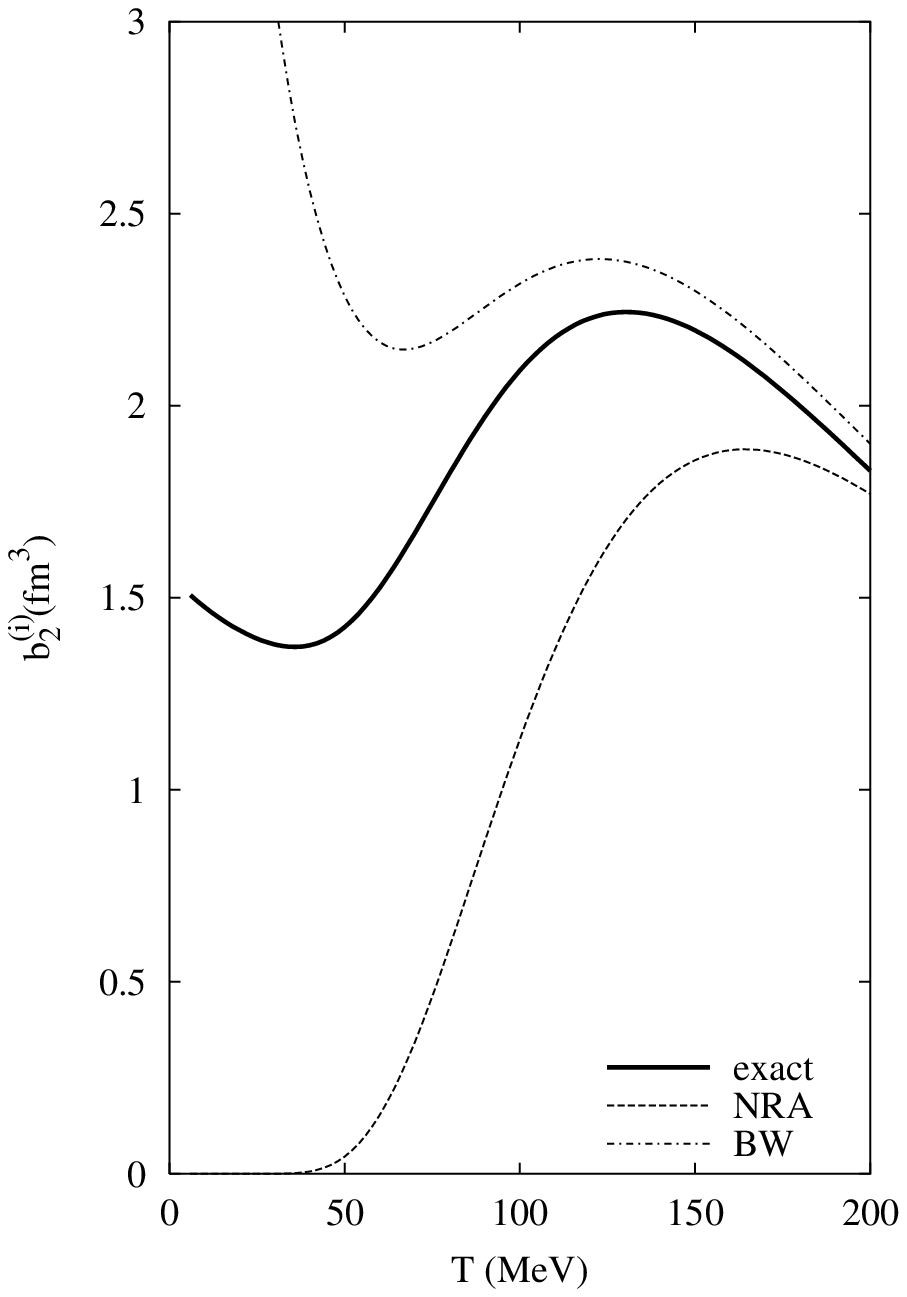}
\vfill
\mbox{}\\
\caption{
The contribution of  $\rho(770)$ to $b_2^{(i)} (T)$.
The exact value is compared to those calculated in narrow 
resonance approximation (NRA) and using normalized Breit-Wigner (BW)
profile of the resonance (\ref{BW-b2}).
\label{frho}
}
\end{center}
\end{figure}

It is easy to see that if a resonance lies far from the threshold:
\begin{equation}\label{BW-crit}
M_r - 2 m >> \Gamma_r 
\end{equation}
both formulae are reduced to
\begin{equation}\label{BW-phase}
 \tan \delta_{r}(q) \approx 
	\frac{\Gamma_r/2} {M_r - \varepsilon}
\end{equation}
In this case the activity of the resonance can be represented in the form
\begin{equation} \label{BW-act}
\sum_r  z_r =   \sum_{I=0}^{2 I_0}  
 {\sum_{L}}^\prime  
\int_{2 m}^{\infty} d \varepsilon 
\zeta(\varepsilon)
 (2I+1) (2L+1)
\phi (T;\varepsilon) \  \exp\left( \frac{2 \mu}{T}  \right),
\end{equation}
where the resonance profile function is given by the Breit-Wigner
formula:
\begin{equation} \label{BW}
\zeta(\varepsilon) =
\frac{1}{2 \pi} \frac{\Gamma_r}
{(\varepsilon - M_r)^2 + (\Gamma_r/2)^2}
\end{equation}
From this we conclude that the procedure of 
Ref.~\cite{Denisenko87} where the profile function was postulated to be 
\begin{equation} \label{BW-dm}
\zeta(\varepsilon) =
 \frac{\xi \Gamma_r}
{(\varepsilon - M_r)^2 + (\Gamma_r/2)^2}
\end{equation}
with normalization constant $\xi$ fixed by the condition
\begin{equation} \label{BW-norm}
\int_{2 m}^{\infty} d \varepsilon  
 \frac{\xi \Gamma_r}{(\varepsilon - M_r)^2 + (\Gamma_r/2)^2} = 1
\end{equation}
becomes valid in the limit (\ref{BW-crit}).
The $\sigma$-resonance obviously does not satisfy this condition,
even for the $\rho(770)$ the difference $M_r - 2 m$ is only about $3$
times larger then the width. We have calculated the contributions
of these two resonances into 
the second cluster integral $b^{(i)}_2$ using the parametrizations 
(\ref{res0}) and (\ref{res}) and compare them with the corresponding 
approximate values found in the framework of the procedure 
of Ref.~\cite{Denisenko87}:
\begin{equation} \label{BW-b2}
  b^{(i)}_2~ =~ \frac{1}{ m^4 T [ g K_2(m/T) ]^2} \sum_r
 \int_0^\infty d q  \varepsilon^2(q)
 \frac{\xi \Gamma_r}
{(\varepsilon - M_r)^2 + (\Gamma_r/2)^2}
 \frac{d \delta_{L I}(q)}{d q}   
 K_2 \left( \frac{\varepsilon(q)}{T} \right)~.
\end{equation}
and in the NRA (\ref{b2NRA}). The results are shown in 
Figs. \ref{fsigma} and \ref{frho}. 
As can be seen from  Fig.\ref{fsigma}, for $\sigma$-resonance 
the both approximations essentially underestimate the CI.
It is interesting to mention that at $T>150$ MeV the 
formula of Ref.~\cite{Denisenko87} gives slightly worse result than even
that of the NRA. In the case of the 
$\rho$-resonance this formula systematically overestimates 
the CI in contrast to the $\sigma$ case. The role of the resonance width
becomes small at high temperatures and all three formulae give comparable
results in both ($\sigma$ and $\rho$) cases.

\section{Interacting pion gas}

\begin{figure}[t]
\begin{center}
\vfill
\leavevmode
\epsfysize=20cm \epsfbox{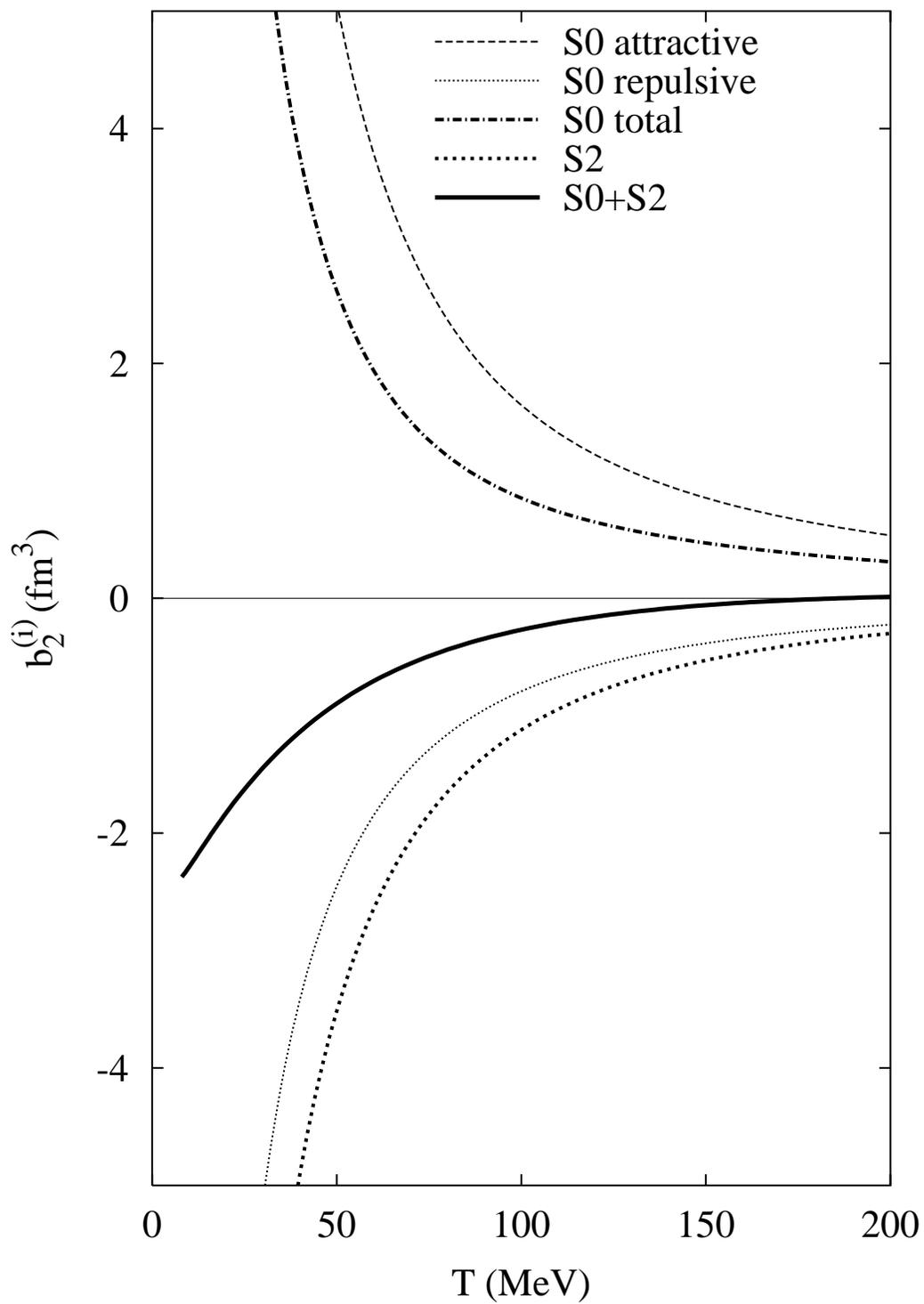}
\vfill
\caption{The contribution of $S_0$ and $S_2$ states 
of $\pi \pi$-scattering into the second CI $b^{(i)}_2$.
\label{fS}
}
\end{center}
\end{figure}

Both type of interaction: hard core repulsion and resonance attraction
are present in the pion gas.
The phase shift for $\pi \pi$-scattering in the $S_0$ state at
the center of mass energy
below $1$ GeV can be represented as a sum of three terms 
\cite{Ishida97,Ishida96,Ishida99}:
\begin{equation}
  \delta_{00}(q) = \delta_{\sigma}(q) + \delta_{f_0}(q) + \delta_{BG}(q) \ .
\end{equation}
The background term $\delta_{BG}$ is related to the hard core 
repulsion
\begin{equation}
  \delta_{BG}(q) = - r_0 q \ , 
	\ \ \ \ \ \ \ \ \ \  
	r_0=3.03 \mbox{ GeV}^{-1}  
\end{equation}
and two first terms describe attraction due to the resonances $\sigma$ and
$f_0$ and are parametrized by the formula (\ref{res0}). 
The contributions of the $S_0$ state to the CI are shown in
Fig.\ref{fS}. 
The attractive part is larger in 
absolute value than the repulsive part, so that total 
contribution of the $S_0$ state is positive.

The interaction in the $S_2$ state has a purely repulsive nature 
(no exotic resonances with isospin $I=2$ have been found).
The phase shift can be successfully fitted by the hard core formula:
\begin{equation}
  \delta_{02}(q) = - r_0 q \ , 
	\ \ \ \ \ \ \ \ \ 
	r_0=0.87 \mbox{ GeV}^{-1} \ .
\end{equation}
As it is seen from Fig.\ref{fS} the absolute value of the negative
contribution of $S_2$ state into the CI is slightly larger than
that of the positive contribution of the $S_0$ state, so that these two
quantities almost cancel each other. The resulting contribution of 
the $S$-state into CI is negative and a few times smaller
than those of the $S_0$- and $S_2$-states separately. 
Due to the small value of the total $S$-state contribution into CI
the $P_1$ state becomes important already at relatively low temperature.

\begin{figure}[t]
\begin{center}
\vfill
\leavevmode
\epsfysize=20cm \epsfbox{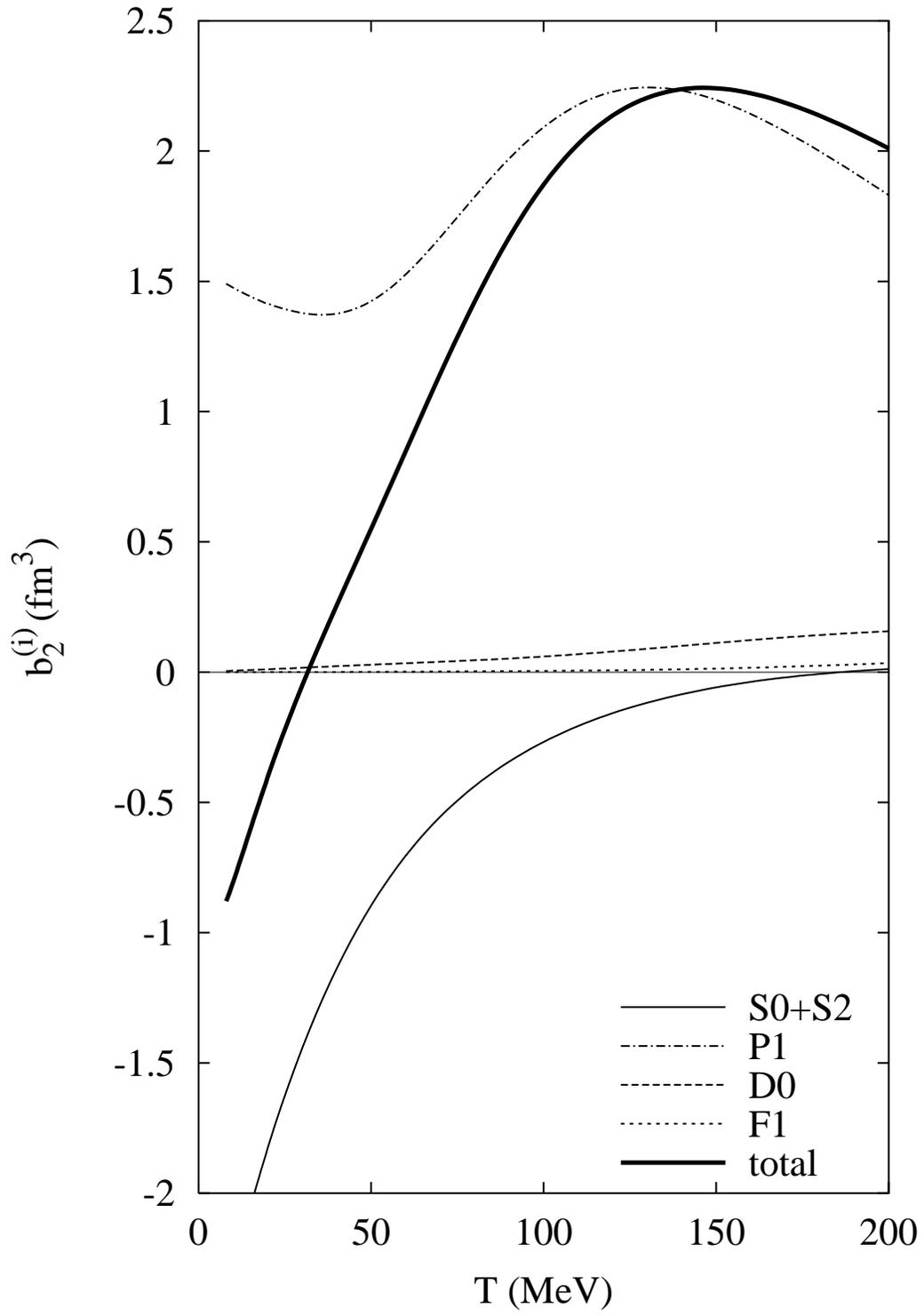}
\vfill
\caption{The partial $\pi \pi$-wave contribution into the second CI 
 $b^{(i)}_2$ and its total value. 
\label{fPW}
}
\end{center}
\end{figure}

The phase shifts of $\pi \pi$-scattering in $P_1$, $D_2$ and $F_1$ states
can be parametrized by the formula (\ref{res}) 
\cite{Hyams73}\footnote{The background 
and higher pole terms which are present in the formulae of
Ref.\cite{Hyams73}
were found to give a negligible contribution to the CI and are dropped
in the present consideration.}.
The results are presented in Fig.\ref{fPW}. At small temperatures 
$T<80$~MeV
both the $S$- and $P$-wave give comparable contributions to the CI. At
higher
temperatures, the $P$-wave dominates. The $D$- and $F$-waves add 
small corrections to the CI at $T>140$~MeV. The contribution of higher
waves
is assumed to be negligible.

It should be mentioned that at very low temperatures, $T < 30$ MeV, the
total
CI becomes negative, in agreement with the results of 
Ref.\cite{Mishustin93} for the isotopically symmetric pion gas, while
 at high temperatures attraction dominates over repulsion.

\begin{figure}[t]
\begin{center}
\vfill
\leavevmode
\epsfysize=20cm \epsfbox{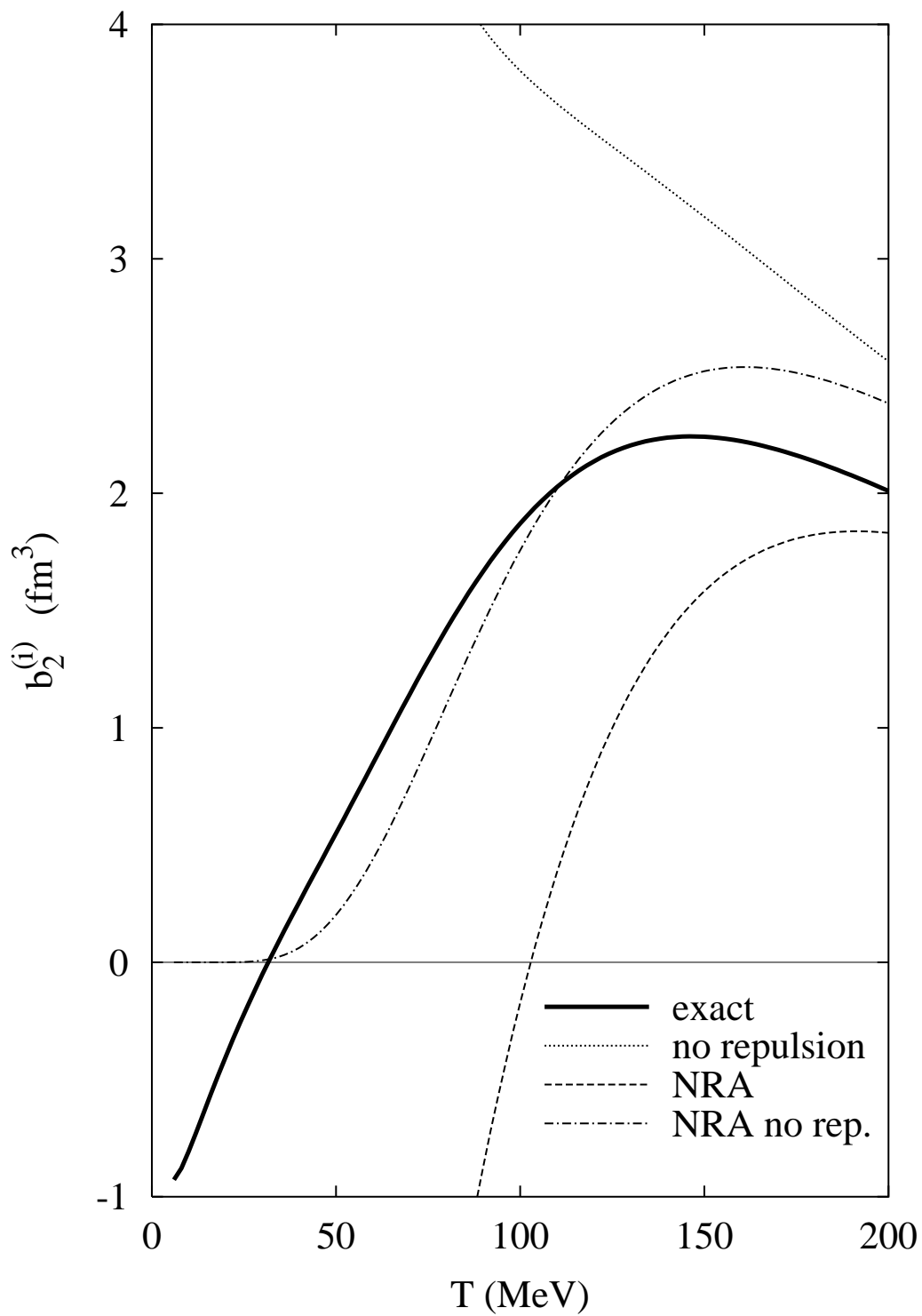}
\vfill
\caption{The exact value of the CI $b^{(i)}_2$ compared to various
approximations: "No repulsion" --- the repulsive part of the 
$\pi \pi$-interaction is dropped out; "NRA" --- narrow resonance 
approximation; "NRA no repulsion" --- both repulsion and final resonance
width are ignored. 
\label{fNR}
}
\end{center}
\end{figure}

 The exact CI is also compared in Fig.\ref{fNR} to various
approximations widely used for the hadron gas analysis. It is seen that 
ignoring repulsion between pions overestimates the CI by more than $35$\%.
On the other hand, the NRA underestimates the CI by at least $20$\%.
At low temperatures $T<120$ MeV both approximations become completely
unreasonable. 
If one ignores both the finite resonance width and the repulsion 
(it corresponds to the ideal gas of pions and 
resonances) these two errors partially cancel each other.
The simplest 
approximation, surprisingly enough, appears to give  
better results than the both
more complicated ones. (There remains, however, discrepancy up to about 
$15$\% at high temperatures).
This means that both effects, the repulsion and the finite resonance
width, should be taken into account simultaneously.
Including either of these effects without another one 
increases rather than decreases the numerical errors with respect to
the simplest IG model of pion and two-pion resonances.

Comparing the CI $b^{(i)}_2$ with the ideal gas CI $b^{(0)}_2$ 
(see Fig. \ref{fdiv})
one observes that the interactions give essential contribution to the CI 
already at $T=70$ MeV. At $T>150$ MeV the interaction part     
$b^{(i)}_2$ clearly dominates over Bose effects related to 
$b^{(0)}_2$.

To estimate the influence of the second CI on the thermodynamical 
properties of the pion gas we have calculated the particle density
in the second cluster approximation
\begin{equation} \label{dens2}
n = n_0 + 2 b^{(i)}_2 z^2, 
\end{equation}
where   
\begin{equation} \label{densPG}
n_0 = \sum_{l=1}^{\infty} l b^{(0)}_l z^l =
\frac{g}{2 \pi^2} 
 \int_{0}^{\infty} d p~ p^2~\frac{1}
 {\exp \left( \frac{\sqrt{p^2+m^2} - \mu}{T} \right) - 1
 }  
\end{equation}
is the density of the ideal pion gas (without resonances).
The calculations were done assuming that the chemical equilibrium,
$\mu=0$, is reached.

\begin{figure}[t]
\begin{center}
\vfill
\leavevmode
\epsfysize=20cm \epsfbox{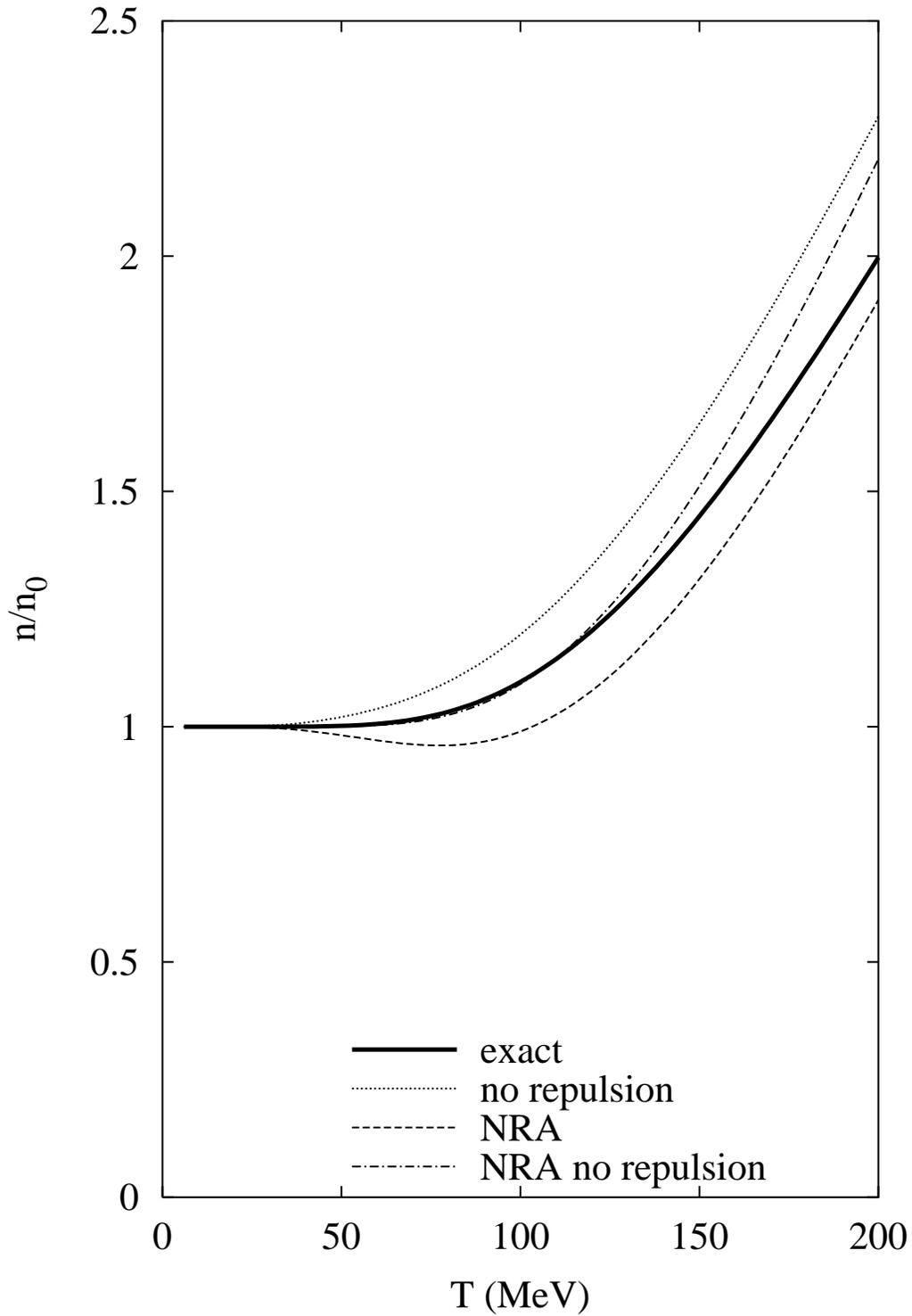}
\vfill
\caption{The ratio of the particle density of the pion gas calculated
in the second order cluster expansion to the particle density of the
ideal pion gas. Various approximations are shown: 
"No repulsion" --- the repulsive part of the 
$\pi \pi$-interaction is dropped out; "NRA" --- narrow resonance 
approximation; "NRA no repulsion" --- both repulsion and final resonance
width are ignored. 
\label{fDEN}
}
\end{center}
\end{figure}

The temperature dependence of the ratio $n/n_0$  
is shown in Fig.\ref{fDEN}. It can be seen that all approximations
give consistent results up to $5 \div 15 \%$. 
A rather small errors is explained by the fact that 
at low temperatures ($T<120$ MeV), where ignoring either the finite
resonance width or the repulsion between pions leads 
to huge errors in the value of cluster integral,  the activity of the
equilibrium pion gas is small and the contribution of the second term of
the cluster expansion into the value of particle density is not important.
On the other hand, at large temperatures, where the second term becomes 
comparable with the first one, the both approximation provide more exact
value of the cluster integral.  
Again, the ideal gas model provides the best approximation at all 
temperatures, except  
very large ones $T > 180$ MeV. 

This conclussion is close to that of Ref.\cite{Venugopalan92}, where it
was pointed out that the interacting pion gas {\it in the second cluster
approximation} only slightly differs from the ideal gas of pions and  
$\rho$-mesons due  to the nearly exact
cancellation of the contributions from S-wave attractive and repulsive 
channels. That is the repulsive interactions and the contribution of the
broad $\sigma$-resonans can be dropped simultaneously.
The aim of our further consideration is to take properly into account the 
hard-core repulsive interactions. In this case the $\sigma$-meson
contribution must be retained.
 
It is seen from Fig.\ref{fDEN} that the contribution of the 
second term of cluster expansion to the particle density is comparable
to that of the ideal gas. There is no reason to expect that the higher 
terms are negligible. The purpose of the next section is to  
go beyond the second cluster approximation.

\section{Van der Waals equation}

\begin{figure}[t]
\begin{center}
\vfill
\leavevmode
\epsfysize=20cm \epsfbox{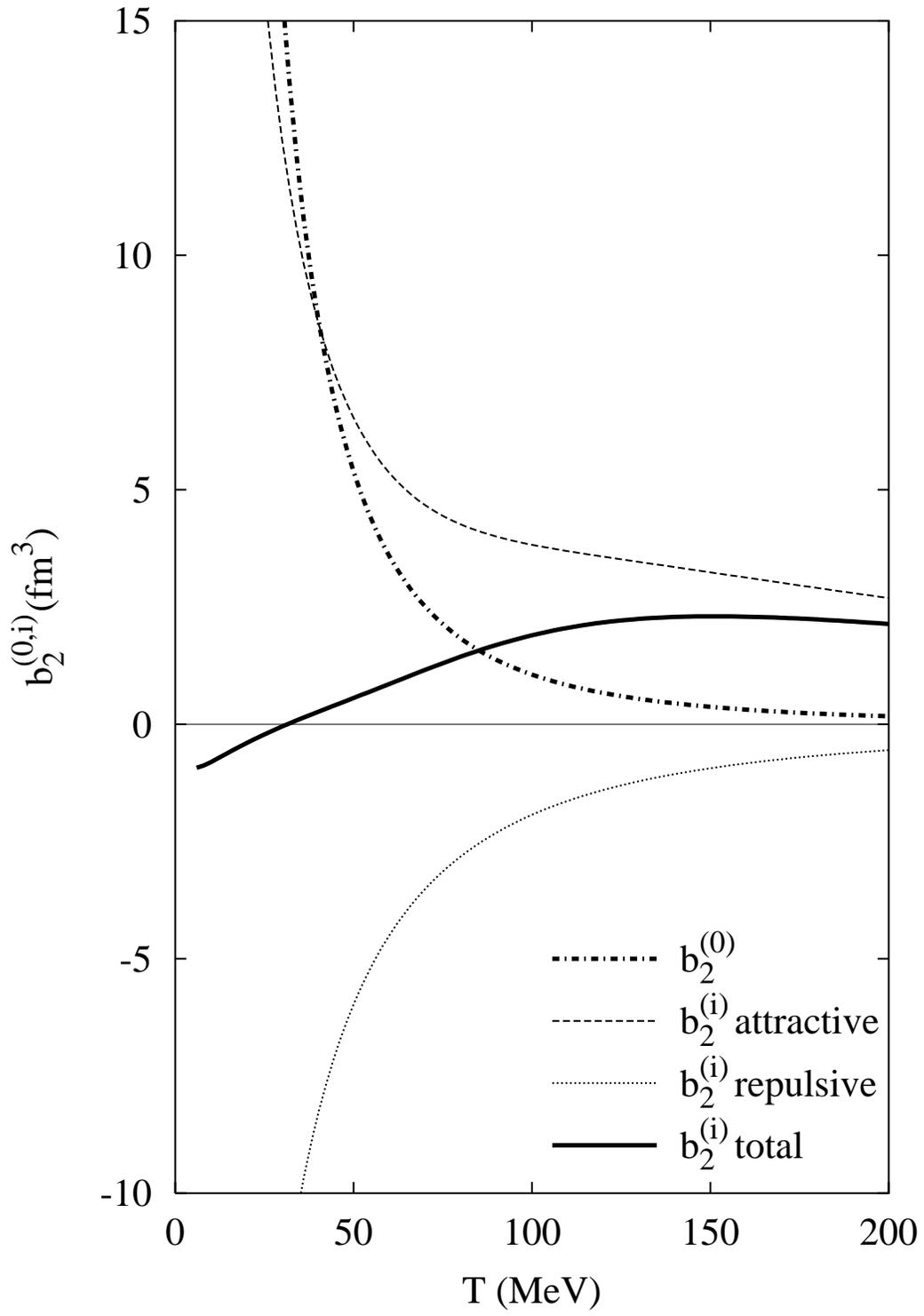}
\vfill
\caption{The attractive and repulsive part of CI 
 $b^{(i)}_2$ and its total value. The ideal gas CI $b^{(0)}_2$ is 
 also shown.
\label{fdiv}
}
\end{center}
\end{figure}

Taking into account the attractive parts of higher cluster terms 
is straightforward: assuming that the attractive interaction of
three and more pions is dominated by resonance interaction (as it
was in the two pion case)  
we just add the activities of all the lightest
pion resonances to the grand canonical partition function logarithm:   
\begin{equation} \label{grandMULT}
 \log {\cal Z}(V,T,\mu) = \log {\cal Z}^{(0)}(V,T,\mu) 
  + 
 V \sum_{r}  z_{r} ,
\end{equation}   
where the first term in the right hand side is the partition function 
of the ideal pion gas
 and the sum in the second term runs over not only
the two pion resonances (Table \ref{respar}),
but also includes the resonances decaying into three and more pions 
(Table \ref{resparM}). For the activity of each resonance species we use 
the following expression
\begin{equation}\label{actMULT}
z_r = 
\int_{N_r m}^{\infty} d \varepsilon 
\zeta(\varepsilon)
g_r \phi (T;\varepsilon) \  \exp\left( \frac{\mu_r}{T}  \right).
\end{equation}
The chemical potential $\mu_r$ of a resonance decaying into $N_r$ pions
is proportional to the pion chemical potential:
\begin{equation} 
\mu_r = N_r \mu
\end{equation}
In our calculations we assume chemical equilibrium: $\mu_r = \mu = 0$.
For the two-pion resonances from Table \ref{respar} we put 
\begin{equation} 
\zeta(\varepsilon) = \frac{1}{\pi} \frac{d \delta}{d \varepsilon} 
\end{equation}
and use the parametrization (\ref{res0}) and (\ref{res}), so that
the contribution of two pion resonances is reduced to the 
$b^{(a)}_2 z^2$, where the $b^{(a)}_2$ is an attractive part of the CI
$b^{(i)}_2$ shown in Fig. \ref{fdiv}.   

For the resonances from Table \ref{resparM} we use the Breit-Wigner
profile function (\ref{BW-dm}) with the normalization (\ref{BW-norm})
\footnote{The formula of Ref.\cite{Denisenko87} does
not lead to large error because of the dominating contribution comes from
the narrow resonance $\omega(782)$, for which both creteria 
(\ref{NRA}) and (\ref{BW-crit}) are fulfilled. On the other hand, lack 
of detailed experimental information on the phase shifts in the vicinity 
on the broad $\pi \rho$- and $\pi \sigma$-resonances as well as large 
uncertainties in their masses, width and decay fractions make impossible
and useless the application of the more exact formula.}.
In the NRA the expression (\ref{actMULT}) is reduced to
Eq.~(\ref{actNRA}).    

It follows from Eq.~(\ref{grandMULT}) that the pressure 
and the pion density are calculated from the ideal gas model for
pion and pion resonances with finite width
(that is the repulsive interactions are ignored): 
\begin{eqnarray}
\label{pMULT}
p(T,\mu)~ &=&~ p_{0}(T,\mu) + T \sum_{r} z_{r} = 
p_{0}(T,\mu) + \sum_{r} p_r(T,\mu_r)               \\
\label{nMULT}
n(T,\mu)~ &=&~ n_{0}(T,\mu) +  \sum_{r} N_r z_{r} =
n_{0}(T,\mu) + \sum_{r} N_r n_r(T,\mu_r),
\end{eqnarray}
where the ideal pion gas pressure $p_0$ is given by the formula
\begin{equation} \label{presPG}
p_0(T,\mu) = 
- g T 
 \int \frac{d^3 p}{(2\pi)^3} \log \left[ 1 - 
 \exp \left( \frac{\mu - \sqrt{{\bf p}^2+m^2}}{T} \right) 
\right],
\end{equation}
with the particle density of the ideal pion gas given by
Eq.~(\ref{densPG}).

To take into account the hard core repulsion between pions and resonances
we use the excluded volume Van der Waals model \cite{Rischke91}.
In the framework of this model the pressure $p^{VdW}(T,\mu)$ of 
one-component gas of particles with the excluded-volume
parameter $v_0$ can be
found from the transcendental equation  
\begin{equation} \label{presVdW1}
p^{VdW}(T,\mu) = p^{id}\left(T,\mu - v_0 p^{VdW}(T,\mu) \right),
\end{equation}
where $p^{id}(T,\mu)$ is the pressure of corresponding ideal gas.
The particle density $n^{VdW}(T,\mu)$ is related to that of ideal gas 
by the expression 
\begin{equation} \label{nVdW1}
n^{VdW}(T,\mu) = 
\frac{ n^{id}\left(T,\mu - v_0 p^{VdW}(T,\mu) \right) }
{1 + v_0 n^{id}\left(T,\mu - v_0 p^{VdW}(T,\mu) \right)},
\end{equation}
The above model can be straightforwardly generalized to a 
multi-component gas, 
if one assumes that all particle species have the same 
excluded-volume parameter.
In our calculations we put it to be the 
same for the pions and pion resonances. 
The standard procedure
of derivation
of the Van der Waals equation in the statistical physics 
shows that the excluded-volume parameter
is equal to the absolute value of
the  
repulsive part of the second virial coefficient \cite{Greiner}.
Therefore, the excluded-volume parameter for the pions can be
identified with the repulsive 
part of the CI $b^{(i)}_2$ (See Fig. \ref{fdiv}):
\begin{equation} 
v_0 = \left|  b^{(r)}_2 \right|~.
\end{equation}
Hence, to find the pressure of the interacting pion gas in the framework of 
the Van der Waals excluded volume model we solve the transcendental equation
\begin{eqnarray}\label{pVdW}
p^{VdW}(T,\mu) &=&  
p_{0}(T,\tilde{\mu}) + \sum_{r} p_r(T,\tilde{\mu}_r)  \\
\tilde{\mu} &=& \mu - v_0 p^{VdW}(T,\mu) \nonumber \ , \\
\tilde{\mu}_r &=& \mu_r - v_0 p^{VdW}(T,\mu) \nonumber \ .
\end{eqnarray}
The particle density of the pions is found from 
\begin{equation} \label{nVdW} 
n^{VdW}(T,\mu)~ =~ \frac{
n_{0}(T,\tilde{\mu}) + \sum_{r} N_r n_r(T,\tilde{\mu}_r) 
}
{ 1 + v_0 \left[
n_{0}(T,\tilde{\mu}) + \sum_{r} n_r(T,\tilde{\mu}_r) 
\right]
}    
\end{equation}

\begin{figure}[t]
\begin{center}
\vfill
\leavevmode
\epsfysize=20cm \epsfbox{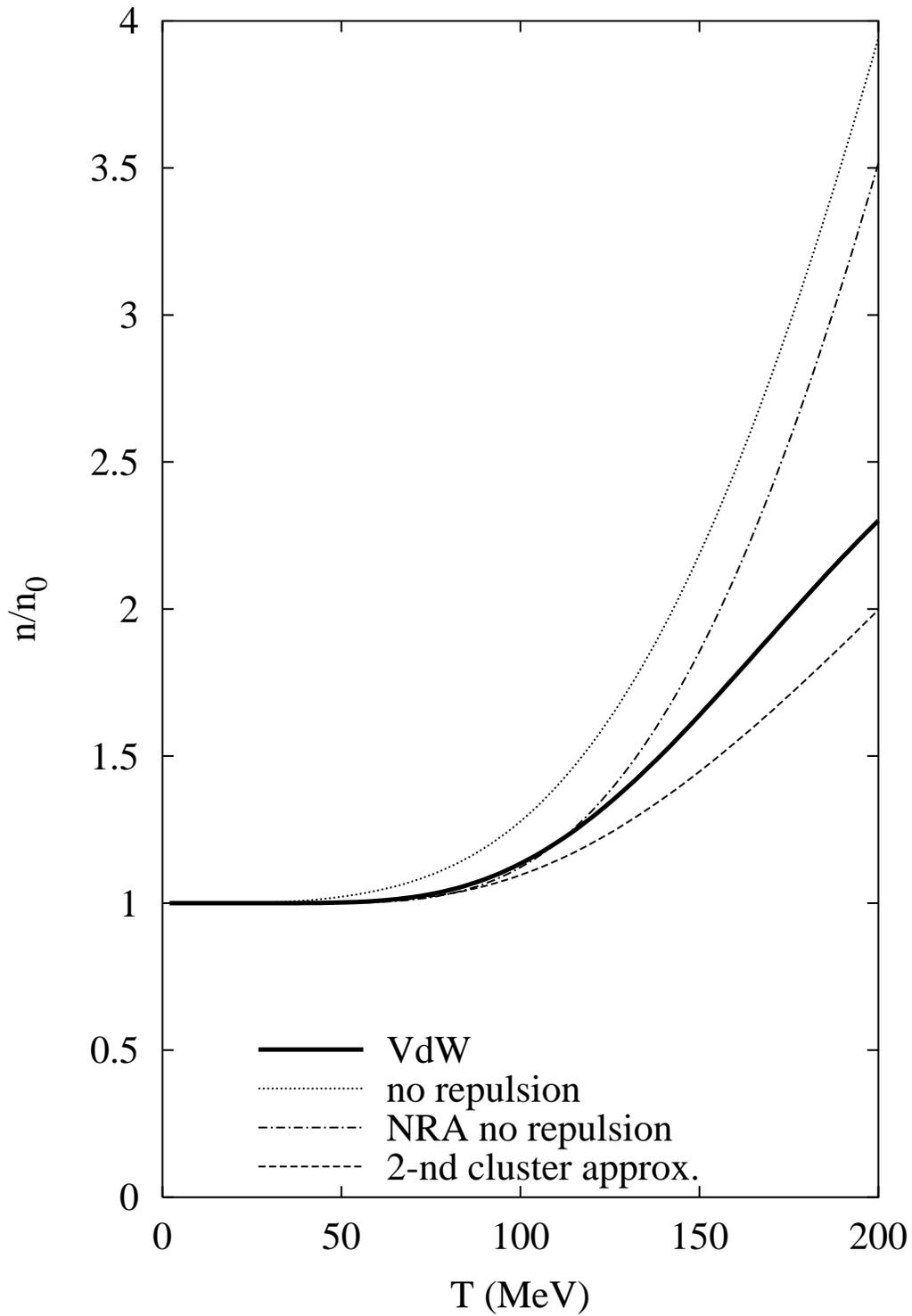}
\vfill
\caption{The pion density in the Van der Waals excluded volume model
compared to the ideal gas of pions and broad pion resonances (no repulsion),
to the ideal gas of pions and zero width pion resonances (NRA no repulsion)
and to the pion gas in the second order cluster approximation.
\label{fMULT}
}
\end{center}
\end{figure}

The result of the calculation is shown in Fig. \ref{fMULT}.
It is seen that essential deviation from the second order cluster 
expansion take place at the temperatures $T \agt 140$ MeV. 
A comparison with the ideal gas model of pions
and pion resonances shows that the effects of the hard-core repulsion
are not cancelled by the effects of 
the finite resonance width. This leads to an essential (up to $30$\%)
suppression of the pion density with respect to the ideal gas 
of pions and pion resonances.

\section{Conclusion}

A quantum mechanical formula for the second 
cluster integral for the gas of relativistic particles with hard-core 
interaction was derived and analyzed. In the nonrelativistic 
classical limit, this formula is reduced to the expression used in 
Refs.\cite{Go:99,Go:97}. In the quantum case, however, the value of the cluster
integral appears to be much larger in magnitude than the corresponding
classical value and, in contrast to the classical case, depends on the 
temperature even in the nonrelativistic limit. It has been demonstrated that 
the second cluster integral for the pion gas all reasonable temperatures is 
far away  from the classical limit. 
Its repulsive part, which can be 
interpreted as  proper particle volume, 
is an order of magnitude 
larger than it could be expected from the classical evaluation. 
It should be mentioned that not only quantum effects but also
relativistic ones are important in the case of pion gas.
Surprisingly, they essentially modify the proper
pion volume even at relatively low temperatures $T \sim 30$ MeV.

The role of finite resonance width in the second cluster integral was 
studied. It was established that the widely used {\it add hoc} formula 
\cite{Denisenko87} with the 
normalized Breit-Wigner resonance profile is unsuitable for 
broad resonances lying close to the threshold, the parametrization of 
the experimental phase shifts should be used instead. 
The most striking example of this kind is the 
$\sigma$-resonance. Our analysis shows that in the case of 
$\sigma$-resonance the calculations with the normalized Breit-Wigner 
profile can give even worse result than simple zero-width
approximation. 

At the second order of cluster expansion, due to the presence of broad 
resonances in the $\pi \pi$-system, the  negative contribution of the 
hard core repulsion into the cluster integral  almost canceled by 
positive contributions of  finite resonance widths in a rather broad 
temperature range. Because of this fact,
the thermodynamical properties of the interacting pion gas in the second 
cluster approximation appear to be quite similar to those of the ideal gas 
of pions and two-pion resonances: the error in the value of the particle 
density does not exceed a few percents. Surprisingly, the account for finite
resonance widths without account for the hard core repulsion as well as 
consideration of the hard core repulsion when the resonance widths are 
neglected worsen rather than improve a simple ideal gas model of pions 
and zero-width pion resonances. Both effects should be either 
neglected or taken into account simultaneously.

% Similar result was obtained in Ref.\cite{Venugopalan92}, 
% where it was found that the contribution of the repulsive interactions and 
% the broad $\sigma$-resonance into the second cluster integral
% almost cancel each other and both can be dropped simultaneously. Similarly, 
% when the hard core repulsion 
% is not neglected, the $\sigma$-resonance must be taken into account,
% and its width cannot be neglected. 

This does not mean, however, that we can restrict ourselves to the simple
ideal gas picture of pions and zero-width pion resonances
at all temperatures. As it has been demonstrated, when the 
temperature is sufficiently high ($T \agt 140$ MeV), the pion density 
becomes so large that the cluster expansion cannot be truncated at the 
second order. In contrast to the second cluster approximation, an appreciable 
deviation from the ideal gas model is observed, when the higher order 
terms are taken into account. 
In the framework of Van der Waals excluded-volume model, the pion density
appears to  be up to $30$\% lower than that of the ideal pion-resonance gas. 
Hence, at high particle densities the correct model of the pion gas 
must include all pion resonances 
and the resonance width as well as the repulsive interactions between the
particles must be taken into account.

It should be emphasized that, if the model takes properly into account 
the hard core repulsion, there is no reason to drop
the $\sigma$-resonance. It must be included into the model along with 
other resonances.

The developed in the present paper approach will be used for calculation 
of excluded volumes of other hadrons. This will allow us to study the 
influence of hard core repulsion on the properties of realistic 
hadron gas including (anti-)nucleons and strange particles
by means of multicomponent Van der Waals equation 
\cite{Gorenstein:1999ce}. We expect essentially larger excluded volume 
effects for nucleons: preliminary 
calculations show that the proper volume of the nucleon is essentially
(by the factor $2 \div 2.5$) larger than that of the pion.   
Thus the hard core repulsion may essentially modify the particle number
ratios in comparison to widely used ideal resonance gas model.

\acknowledgements
We thank K.Bugaev for helpful discussions and comments.
We acknowledge the financial support of DAAD and DFG, Germany.
The research described in this publication was made possible in part by
Award No. UP1-2119 of the U.S. Civilian Research \& Development
Foundation for the Independent States of the Former Soviet Union
(CRDF).

\begin{table}
\caption {Parameters for the lightest resonances in  
$\pi \pi$-system.
\label{respar}
}
\begin{tabular}{|c|c|c|c|c|c|c|}
Resonance & Isospin & Spin & Mass &
Width  & Elasticity & Interaction  \\ 
       & $I_r$ & $L_r$ & $M_r$ (MeV) &
$\Gamma_r$ (MeV) & $x_r$ & radius $R_r$ (GeV${}^{-1}$)\\ \hline
$\sigma$ & $0$ & $0$ &  $585$  &  $385$  & --- & --- \\
$f_0(980)$     & $0$ & $0$ & $993.2$ & $54.32$ & --- & --- \\
$\rho(770)$    & $1$ & $1$ &  $777$  &  $155$  & $1$ & $3.09$ \\
$f_2(1270)$    & $0$ & $2$ & $1281$  &  $205$  & $0.84$ & $4.94$ \\
$\rho_3(1690)$ & $1$ & $3$ & $1713$  &  $228$  & $0.26$ & $6.38$ \\
\end{tabular}
\end{table}

\begin{table}
\caption {Parameters for the lightest resonances decaying into $3$ and
more pions.
\label{resparM}
}
\begin{tabular}{|c|c|c|c|c|c|}
Resonance    & Degeneration &  Mass  & Width      & Elasticity & Number of 
                                                                 pions in  \\ 
             & factor $g_r$ & $M_r$ 
						                   (MeV) & $\Gamma_r$ 
															           (MeV)    & $x_r$      & the final 
																				                         state 
																				                       $N_{r}$\\ 
\hline
$\omega(782)$&    $3$       & $782$  &  8.41      &  0.888     & $3$ \\   
$\phi(1020)$ &    $3$       & $1019$ &  4.43      &  0.155     & $3$ \\
$h_1(1170)$  &    $3$       & $1170$ &  360       & $\sim 0.5$ & $3$ \\ 
$b_1(1235)$  &    $9$       & $1230$ &  142       & $\sim 1 $  & $4$ \\
$a_1(1260)$  &    $9$       & $1230$ &  425       & $\sim 1 $  & $3$ \\
$f_1(1285)$  &    $3$       & $1282$ &  24        &  $0.35$    & $4$ \\   
$\pi(1300)$  &    $3$       & $1300$ &  400       & $\sim 1 $  & $3$ \\
$a_2(1320)$  &   $15$       & $1318$ &  107       & $0.70$     & $3$ \\
\end{tabular}
\end{table}

\end{document}